\documentclass[aps,prb,twocolumn,groupedaddress]{revtex4}

\usepackage{graphicx}

\begin{document}

\title{High resolution measurements of the switching current in a
Josephson tunnel junction: Thermal activation and macroscopic
quantum tunneling}

\author{A. Wallraff}
\email[]{wallraff@physik.uni-erlangen.de}
\affiliation{Physikalisches Institut III, Universit{\"a}t
Erlangen-N{\"u}rnberg, D-91058 Erlangen, Germany}
\author{A. Lukashenko}
\altaffiliation{Scientific Centre of Physical Technologies,
National Academy of Sciences, 61145 Kharkov, Ukraine}
\author{C. Coqui}
\affiliation{Physikalisches Institut III, Universit{\"a}t
Erlangen-N{\"u}rnberg, D-91058 Erlangen, Germany}
\author{T. Duty}
\altaffiliation[Previous address: ]{D-Wave Systems Inc.  1985 West
Broadway Vancouver, BC V6J 4Y3, Canada}
\altaffiliation[Current address: ]{Department of Microelectronics and
Nanoscience, MC2, Chalmers University of Technology and G{\"o}teborg
University, S-412 96 Gothenburg, Sweden}
\author{A.~V. Ustinov}
\affiliation{Physikalisches Institut III, Universit{\"a}t
Erlangen-N{\"u}rnberg, D-91058 Erlangen, Germany}

\date{\today}

\begin{abstract}
    We have developed a scheme for a high resolution measurement of the
    switching current distribution of a current biased Josephson tunnel
    junction using a timing technique.  The measurement setup is
    implemented such that the digital control and read-out electronics are
    optically decoupled from the analog bias electronics attached to the
    sample.  We have successfully used this technique to measure the
    thermal activation and the macroscopic quantum tunneling of the phase
    in a small Josephson tunnel junction with a high experimental
    resolution.  This technique may be employed to characterize
    current-biased Josephson tunnel junctions for applications in quantum
    information processing.
\end{abstract}


\maketitle

\section{Introduction}

The current-biased Josephson tunnel junction is an ideal system to
study both thermal activation \cite{Kramers40,Haenggi90b} and quantum
tunneling\cite{Calderia81,Devoret92} in a controllable experimental
environment.  The dynamics of such a junction is equivalent to that of
a particle in a tilted washboard potential\cite{Stewart68,McCumber68}. 
The process of the particle escape from a metastable state in this
system can be characterized, by analyzing the transition of the
particle from a state in which it is localized in a potential well to
a state in which it runs down the potential.  This corresponds to the
transition of the Josephson junction from the superconducting
zero-voltage state to a finite voltage state in the presence of an
applied bias current.  At high temperatures the escape is dominated by
thermal activation across the barrier, at low temperatures it is
determined by quantum tunneling through the barrier.  The rate with
which the particle escapes from the well depends on the detailed shape
of the potential, the dissipation in the system and the temperature of
the thermal bath to which the system is coupled.

Thermal activation in a current biased Josephson junction has been
studied both theoretically and experimentally for large damping
\cite{Martinis89,Vion96}, intermediate to low damping
\cite{Fulton74,Castellano96b,Ruggiero98} and extremely low damping
\cite{Martinis87,Ruggiero98}.  At low temperatures the quantum
mechanical properties of Josephson junctions have been investigated in
relation to macroscopic quantum tunneling (MQT)
\cite{Jackel81,Voss81,Devoret85}, energy level quantization (ELQ)
\cite{Martinis87,Silvestrini97} and macroscopic quantum coherence (MQC)
\cite{Han01}.  Recently, the quantum mechanical properties of
Josephson junction systems have regained interest in the view of their
possible application for solid state based quantum information processing
\cite{Bocko97,Mooij99a,Ioffe99}.  The prospects of superconducting
devices containing Josephson junctions for use as carriers of
quantum information has been strengthened by recent encouraging
experimental results \cite{Nakamura99,Friedman00,vanderWal00}.

In this paper we present a measurement technique implemented for the
characterization of the quantum properties of current biased phase
qubits.  The measurement scheme is based on a high resolution
measurement of the junctions switching current \cite{Fulton74} using a
timing technique.  The measurement setup was tested by performing
measurements of both thermal activation and quantum tunneling in a
current biased small Josephson junction.

In Sec.~\ref{sec:model}, the Stewart-McCumber model describing the
dynamics of the phase in a Josephson junction is briefly reviewed in
relation to thermal activation and quantum tunneling of the phase. 
The measurement technique and the setup implemented for its
realization is described in Sec.~\ref{sec:technique}.  The switching
current distributions obtained over a large range of temperatures are
presented in Sec.~\ref{sec:exp}.  The relative resolution of these
measurements is compared with other results from literature in
Sec.~\ref{sec:res}.  In Sec.~\ref{sec:analysis}, the data in the
thermal regime is analyzed in the view of the low dissipation in the
junction.  The crossover temperature to the quantum regime is
determined and the measured switching current distributions are
compared to the quantum predictions.  Finally the results are
summarized in Sec.~\ref{sec:conclusions}.

\section{Model \label{sec:model}}
\begin{figure}[tb]
 	\centering
	\includegraphics[width = 1.0\columnwidth]{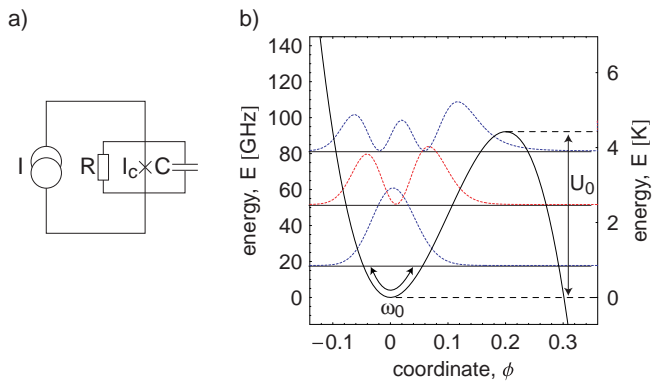}
	\caption{a) Resitively, capacitively shunted junction (RCSJ) model.
	b) $U^{\phi}$ versus $\phi$ calculated for a junction with the
	parameters as used in experiments and biased at $\gamma = 0.995$.  The
	barrier height $U^{\phi}_{0}$ and the oscillation frequency
	$\omega^{\phi}_{0}$ are indicated.  Numerically calculated energy
	levels and the squared wave functions are shown.}
	\label{fig:potentialQmG995}
\end{figure}

In the Stewart-McCumber model \cite{Stewart68,McCumber68}, the
dynamics of a current biased Josephson tunnel junction is described by
an equation of motion for the phase difference $\phi$
\begin{equation}
    C \left(\frac{\Phi_{0}}{2\pi}\right)^{2}
    \ddot{\phi} + \frac{1}{R} \left(\frac{\Phi_{0}}{2\pi}\right)^{2}
    \dot{\phi} + I_{c} \frac{\Phi_{0}}{2\pi} \sin({\phi}) - I
    \frac{\Phi_{0}}{2\pi} = 0\, ,
    \label{eq:smalljunctiondynamics}
\end{equation}
where $\Phi_{0}$ is the magnetic flux quantum, $C$ is the capacitance,
$R$ the effective junction resistance and $I_{c}$ the fluctuation free
critical current of the junction, see Fig.~\ref{fig:potentialQmG995}a. 
Here, $I$ is the externally applied bias current.  The equation of
motion (\ref{eq:smalljunctiondynamics}) is equivalent to the damped
motion of a particle of mass $m_{\phi} = C (\Phi_{0}/2\pi)^{2}$ in an
external potential $U^{\phi}(\phi)$ along the generalized coordinate
$\phi$
\begin{equation}
    m_{\phi} \ddot{\phi}
    +  m_{\phi} \frac{1}{R C} \dot{\phi}
    +  \frac{\partial U^{\phi}(\phi)}{\partial \phi} = 0\, .
    \label{eq:smalljunctiondynamicsparticle}
\end{equation}
In Eq.~(\ref{eq:smalljunctiondynamicsparticle}) the damping 
coefficient is $1/RC$ and the potential is given by
\begin{eqnarray}
	U^{\phi}(\phi) & = & E_{J}\left(-\gamma
	\phi -\cos \phi  \right), 
	\label{eq:potSJ}
\end{eqnarray}
where $E_{J} = \Phi_{0}I_{c}/2\pi$ is the Josephson coupling energy
and $\gamma = I/I_{c}$ is the normalized bias current. 
$U^{\phi}(\phi)$ is a cosinusoidal potential (\ref{eq:potSJ}) with an
amplitude proportional to $E_{J}$, which is tilted proportionally to the
applied bias current $\gamma$.  Because of these properties
$U^{\phi}(\phi)$ is called often washboard potential.

In the absence of thermal and quantum fluctuation and for bias currents
$\gamma < 1$, the junction is in the zero voltage state, corresponding
to the particle being localized in one of the potential wells (see
Fig.~\ref{fig:potentialQmG995}b).  At finite temperatures $T > 0$, the
particle may escape from the well at bias currents $\gamma < 1$ by
thermally activated processes
\cite{Fulton74,Silvestrini88b,Silvestrini88,Turlot89,Vion96,Castellano96b}
and also by quantum tunneling through the barrier
\cite{Voss81,Jackel81,Washburn84,Devoret85,Martinis87}.  The rate at
which both processes occur depends on the
barrier height 
\begin{eqnarray}
    U^{\phi}_{0} & = & 2 E_{J} \left[\sqrt{1-\gamma^2} - \gamma
    \arccos(\gamma) \right] \, ,
\end{eqnarray}
the oscillation frequency of the particle at the bottom of the well
\begin{equation}
    \omega^{\phi}_{0} = \sqrt{U''^{\phi}/m_{\phi}} = \omega_{p}
    \left(1-\gamma^{2}\right)^{1/4} \, ,
    \label{eq:smallFreqSJ}
\end{equation}
and the damping in the junction.  Here $\omega_{p} = \sqrt{2 \pi I_{c}
/ \Phi_{0} C}$ is the plasma frequency. For $\gamma \rightarrow 1$, 
Eq.~(\ref{eq:smallFreqSJ}) can be approximated as
\begin{eqnarray}    
    U^{\phi}_{0} &\approx& E_{J} \frac{4\sqrt{2}}{3} 
    (1-\gamma)^{3/2} \, .
    \label{eq:VoSJ}
\end{eqnarray}

At high temperatures the escape of the particle from the well is
dominated by thermally activated processes, which occur with a
bias-current dependent rate of \cite{Kramers40,Haenggi90b}
\begin{equation}
	\Gamma_{\rm{t}} = a_{t} \frac{\omega^{\phi}_{0}}{2\pi} 
	\exp\left(-\frac{U^{\phi}_{0}}{k_{\rm{B}} T}\right) \, .
     \label{eq:thermalrate}
\end{equation}
Here $a_{t}$ is a temperature and damping dependent thermal prefactor.

As the temperature $T$ is lowered, the thermal activation is
exponentially suppressed.  At temperatures below the so called
cross-over temperature $T^{\star}$ the quantum tunneling rate of the
particle through the barrier exceeds the thermal activation
rate\cite{Grabert84}.  The tunneling rate of the phase is calculated
in the Wentzel-Kramers-Brillouin (WKB) approximation for a particle of
mass $m^{\phi}$ tunneling through the potential barrier described by
$U^{\phi}$.  The effect of damping with the characteristic coefficient
$a = 1/ (2 R C \omega^{\phi}_{0}) = 1/(2Q)$, where $Q =
\omega_{0}^{\phi} RC$ is the junction quality factor, is considered in
terms of a coupling to a bath of harmonic oscillators
\cite{Calderia81,Caldeira83,Leggett84}.  The rate is then given by
\begin{equation}
	\Gamma_{\rm{q}} = A \exp(-B)
	\label{eq:WKBrate}
\end{equation}
with
\begin{eqnarray}
	A & = & \sqrt{60} 
	\omega^{\phi}_{0}\left(\frac{B}{2\pi}\right)^{1/2} (1+{\cal O}(a)),
	\label{eq:coeffA} \\
	B & = & \frac{36 U^{\phi}_{0}}{5 \hbar \omega^{\phi}_{0}} 
	(1+1.74a+{\cal O}(a^{2})).
	\label{eq:coeffB}
\end{eqnarray}
As in Eq.~(\ref{eq:thermalrate}), the values of $\Gamma_{q}$,
$U_{0}^{\phi}$, and $\omega_{0}^{\phi}$ depend on $I$.

At low temperature and small damping the energy of the small
oscillations of the phase at the bottom of the well is quantized, see
Fig.~\ref{fig:potentialQmG995}b.  The energy level quantization has
been observed experimentally both below \cite{Martinis87} and above
$T^{\star}$ \cite{Silvestrini97,Ruggiero99a}.  The escape of the phase
from the well in the presence of quantized energy levels was discussed
theoretically by Larkin and Ovchinnikov in Ref.~\onlinecite{Larkin86}. 
This theory allows for the calculation of the bias current dependent
escape rate $\Gamma(I)$ considering the escape of the particle from
the well by tunneling from any energy level.  The occupation of the
levels can be calculated for finite temperatures and in the presence
of microwaves using a master equation approach.  Previous measurements
of energy level quantization have been compared with the predictions
of this theory and good agreement was found\cite{Kopietz88,Chow88}. 
This theory has also been used to explain switching current
distributions for non-stationary distributions of the phase at high
current ramp rates and above the cross-over temperature
\cite{Ruggiero99a}.  Here, we analyze our data for the escape of the
phase in the presence of energy levels with a separation comparable to
the bath temperature $k_{\rm{B}} T$ at temperatures above $T^{\star}$ in
the framework of this theory.


\section{Measurement Technique and Setup \label{sec:technique}}
The escape of the phase in a Josephson tunnel junction is
experimentally investigated by performing a statistical measurement of
the current at which the junction switches from the zero-voltage state
to a finite voltage state\cite{Fulton74}.  There are two different
well established methods to perform such a measurement.  One method
consists of applying a fixed bias current $I<I_{c}$ to the junction
and measuring the time between the application of the current and the
appearance of a voltage across the junction \cite{Devoret84,Han01}.  This
corresponds to a direct measurement of the lifetime $\tau(I)$ of the
particle in the well.  The inverse of the lifetime corresponds to the
escape rate $\Gamma(I)$.  In the other method, the bias current
applied to the junction is ramped up at a constant rate $\dot{I}$ and
the current $I$ at which the junction switches from the zero-voltage
to a finite voltage state is recorded \cite{Fulton74}.  The switching
current probability distribution $P(I)$ is found by accumulating a
large number of switching currents $I$ and generating a histogram. 
From the $P(I)$ distribution the bias current dependent escape rate
\begin{equation}
    \Gamma(I) = \dot{I} \ln{\frac{\int_{I}^{\infty}P(I')dI'}
    {\int_{I + \Delta I}^{\infty}P(I')dI'}} \, .
    \label{eq:RateCont}
\end{equation}
can be reconstructed \cite{Fulton74} and compared with the theoretical
predictions.

\subsection{Experimental setup}
In order to measure the switching current distribution of a Josephson
tunnel junction with a high accuracy, we have developed a measurement
setup in which the analog biasing electronics and the sample are
electrically isolated from the digital control electronics using an
optical fiber link.  Using a universal time interval counter with $20
\, \rm{GHz}$ time-base, the switching current is determined by
measuring the time between the zero crossing of the bias current
ramped up at a constant rate and the switching of the junction to the
finite voltage state.

\begin{figure*}[tb]
 	\centering
	\includegraphics[width = 0.9 \textwidth]{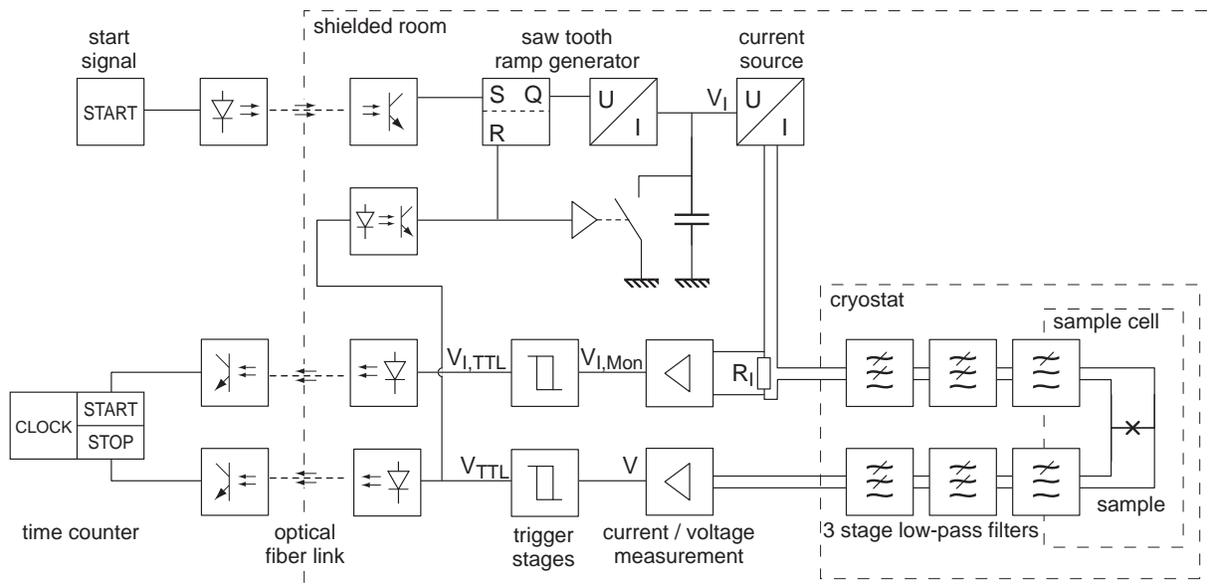}
	\caption{Schematic of the measurement setup. (A detailed explanation 
	is found in the text.)}
	\label{fig:AnalogElectronics}
\end{figure*}

In the following the elements of the experimental setup are discussed
starting from the sample cell mounted at the cold finger of the
dilution refrigerator, the cold and warm filters in the dc-bias lines,
the room temperature analog electronics, the optical fiber link and
finally the digital control and data acquisition electronics, see 
Fig.~\ref{fig:AnalogElectronics}.

\subsubsection{Sample Cell and Wiring}
The sample, typically fabricated on a $5 \times 5 \, \rm{mm^2}$ chip,
is mounted in a closed rf-tight copper cell to shield it from
electromagnetic radiation from the environment.  The sample cell is
thermally anchored to the cold finger of a dilution refrigerator.  For
current-biasing the Josephson junction and measuring the voltage
across it, four wires are fed through microwave filters into the cell. 
Inside the cell the wires are thermally anchored and connected via
wire bonds to the sample.  Each microwave filter consists of a
commercially available \footnote{Thermocoax GmbH, Hamburg, Germany} 1 meter
piece of a $0.5 \, \rm{mm}$ in diameter, lossy stainless steel coaxial line
with a $50 \, \rm{\Omega/m}$ inner conductor made from Ni/Cr
(80\%/20\%) which is soldered into the wall of the cell.  This type of
filter, also called thermocoax filter \cite{zorin95}, has an
attenuation of more than $50 \, \rm{dB}/m$ at frequencies above $1 \,
\rm{GHz}$.  Alternatively, we have also used copper powder microwave
filters \cite{Martinis87}, consisting of a few meters of resistive
wire coiled up in a $10 \, \rm{cm}$ long copper tube filled with $10
\, \rm{\mu m}$ grain-size copper powder and mounted in the wall of the
cell.  If heating of the sample at large bias currents is an issue,
the total dc resistance of this filter may easily be adjusted by
choosing a different wire material and diameter.  This type of filter
is designed to have an attenuation of more than $50 \, \rm{dB}$ per
piece above $1 \, \rm{GHz}$.  All required additional wires for
heaters, temperature sensors, coils etc.~were fed into the cell
through microwave filters of this type.  Another possible type of
cryogenic microwave filter, which has not been tested in our setup,
is based on a micro-fabricated distributed RLC filter \cite{Vion95}.

The sample leads are passed to the 1 K stage of the cryostat in a
tightly twisted loom of superconducting wire and filtered using a low
pass RC filtering stage with a $3 \, \rm{dB}$ cutoff frequency of
approximately $50 \, \rm{kHz}$, which is thermally anchored at the $1
\, \rm{K}$ pot.  The wiring is further fed to the top of the cryostat
in a loom of twisted copper pairs shielded separately from the sensor and
control wiring of the cryostat.  At room temperature all wires are
going through low pass $\pi$-type feedthrough filters with a cutoff
frequency of about $10 \, \rm{MHz}$.  After the last filtering stage
the wires are separated into shielded pairs which are connected to the
analog biasing electronics.

\subsubsection{Analog Electronics}
The sample is current biased by a voltage controlled current source with
selectable output-current ranges.  The source current $I$ is
proportional to the input voltage $V_{I}$ in the range from $-10 \,
\rm{V}$ to $10 \, \rm{V}$.  The control voltage $V_{I}$ is a sawtooth
waveform with an adjustable voltage ramp rate generated by charging a
capacitor with a constant current.  The sawtooth signal is started
upon a digital trigger signal at the set input (S).  The voltage ramp is
stopped by a digital trigger signal at the reset input (R), see 
Fig.~\ref{fig:AnalogElectronics}.

The current flowing through the sample is monitored by measuring the
voltage drop $V_{I,\rm{Mon}}$ across the biasing resistor $R_{I}$
using an instrumentation amplifier.  The voltage $V$ across the
junction is measured using a fast FET instrumentation amplifier
(amplification $\times 1000$).  Both voltages $V_{I,\rm{Mon}}$ and $V$
are fed to two independent Schmidt triggers with adjustable threshold
and window voltages.  The current trigger is calibrated to generate a
TTL pulse $V_{I,\rm{TTL}}$ upon the zero-crossing of the bias current. 
The voltage trigger is set up to generate a TTL pulse $V_{\rm{TTL}}$
when the junction switches from the zero voltage to a finite voltage
state.  The trigger window is adjusted wide enough to avoid voltage
noise induced triggering.  The voltage trigger signal $V_{\rm{TTL}}$
is simultaneously used to stop the current ramp.  To avoid a ground
loop in the signal lines an opto-coupler is used at the reset input of
the sawtooth generator.

All analog electronics components described above have been
specifically designed and implemented for these measurements.  The
supply voltage is delivered by lead accumulators.  Care has been taken
to avoid ground loops.  All electronics is grounded at a single
grounding point on the top of the cryostat.  The cryostat and the
analog electronics are placed inside a stainless steel shielded
room.  Outside of the shielded room the pumping lines of the cryostat
are electrically isolated from the gas handling system and the pumps. 
The cryostat is connected to the shielded room through the pumping
lines.

To perform the current measurement, the current trigger signal
$V_{I,\rm{TTL}}$ and the voltage trigger signal $V_{\rm{TTL}}$ are
converted to digital optical pulses which are passed out of the
shielded room via optical fibers.  Outside the shielded room both
signals are converted back to TTL level electric signals.  Conversely,
for starting the current ramp a TTL pulse is passed via the optical
fiber link into the shielded room.  The arrangement of the analog
electronics setup is shown in Fig.~\ref{fig:AnalogElectronics}.

\subsubsection{Digital Control and Data Acquisition Electronics}

A square wave generator provides a TTL signal at a rate $\nu_{I}$
which is used to start the current ramp with the same repetition
rate $\nu_{I}$.  In each cycle of the measurement the bias current is
increased at a fixed rate $\dot{I}$ until the switching current $I$ is
reached.  The optically decoupled current and voltage trigger signals
are then supplied to the start and stop inputs of a universal time
interval counter to measure the time $\tau$ between the triggers with
a resolution of approximately $25 \, \rm{ps}$.  Having calibrated the
current ramp rate $\dot{I}$ the switching current can be calculated
from the time as $I = \dot{I} \tau$.

Our method has a high instrumental resolution $\Delta I = \dot{I} \,
25 \,\rm{ps}$ of the current measurement.  For typical ramp rates
between $0.1 \, \rm{A/s}$ and $1 \, \rm{A/s}$ as used in the
experiments presented here, a relative \emph{instrumental current
resolution} between $\Delta I / I = 2.5 \, 10^{-12}$ and $25.  \,
10^{-12}$ can be achieved.  This value is much better than that
achievable using a 16 bit A/D converter for direct current
measurements with $\Delta I / I = 0.15 \, 10^{-6}$.

Similar time-based methods for measuring the junctions critical
current at a constant current ramp rate
\cite{Silvestrini88,Silvestrini88b} or the lifetime of the
zero-voltage state of a junction at a constant applied bias
current \cite{Devoret84,Han01} have been implemented by other groups.

The time interval counter \footnote{Stanford Research Instruments,
Model SR 620 Universal Time Interval Counter} used in our setup allows
for ``the on-the-fly'' generation of switching current histograms from
the acquired data during the measurement.  This feature is very useful
for online monitoring of switching current histograms with changing
experimental conditions, e.g. when doing spectroscopic measurements by
applying microwave radiation to the sample or when searching for
parasitic noise sources.

The temperature controller of the dilution refrigerator is installed
inside the shielded room.  It can be powered either by two lead
accumulators or by an isolation transformer.  The electrical ground of
the controller is anchored on the cryostat.  The temperature
controller is computer controlled via an optical fiber link.

\section{Experimental Observations \label{sec:exp}}

The experiments were performed using a high quality $5 \times 5 \,
\rm{\mu m^2}$ tunnel junction fabricated on an oxidized silicon waver
using a standard Nb/Al-AlO$_{x}$/Nb trilayer process.  The junction
had a nominal critical current density of $j_{c} \approx 1000 \,
\rm{A/cm^2}$, a nominal capacitance of $C \approx 1 \, \rm{pF}$ and a
subgap resistance of $R > 500 \, \rm{\Omega}$ below $1.0 \, \rm{K}$.

\begin{figure}[tb]
 	\centering
	\includegraphics[width = 0.95 \columnwidth]{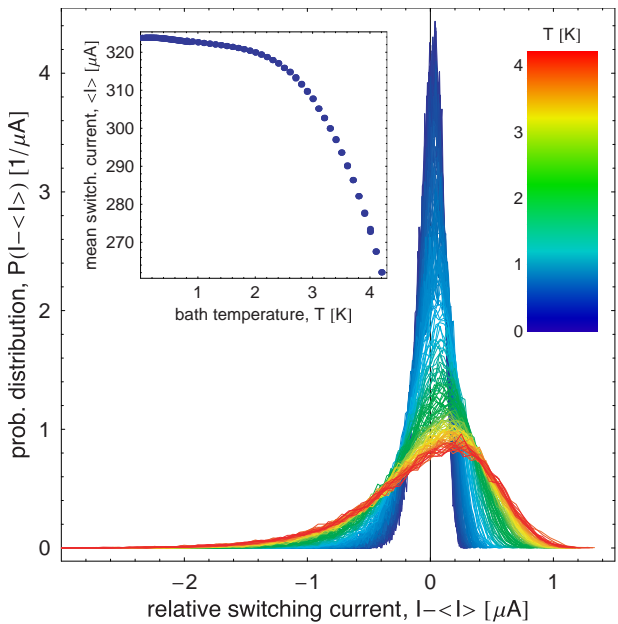}
	\caption{Switching current distributions $P(I-\langle I \rangle)$ at
	bath temperatures between $T = 4.2 \, \rm{K}$ and $T = 25 \, \rm{mK}$. 
	The temperature is color coded according to the scale shown in the 
	plot. 
	In the inset the mean value $\langle I \rangle$ of
	the switching current distribution is plotted versus the bath
	temperature $T$.}
	\label{fig:P(I)summary}
\end{figure}

The switching current distribution of the sample was measured at bath
temperatures $T$ between $4.2 \, \rm{K}$ and $25 \, \rm{mK}$.  The
current was ramped up at a constant rate of $\dot{I} = 0.245 \,
\rm{A/s}$ with a repetition rate of $\nu_{I} = 500 \, \rm{Hz}$. 
Typically $10^{4}$ switching currents were recorded at each
temperature.  Histograms of the switching currents were calculated
with bin widths of approximately $10 \, \rm{nA}$ in order to determine
the $P(I)$ distributions.  In Fig.~\ref{fig:P(I)summary} the measured
switching current distributions are plotted versus $T$.  To show all
histograms in the same plot, the switching probability $P$ has been
plotted versus $I - \langle I \rangle$, where $\langle I \rangle$ is
the temperature dependent mean value of the switching current shown in
the inset of Fig.~\ref{fig:P(I)summary}.  It is observed that the
width of the $P(I)$ distributions decreases with temperature and then
saturates at low temperatures, as expected.

\begin{figure}[tb]
 	\centering
	\includegraphics[width = 0.95 \columnwidth]{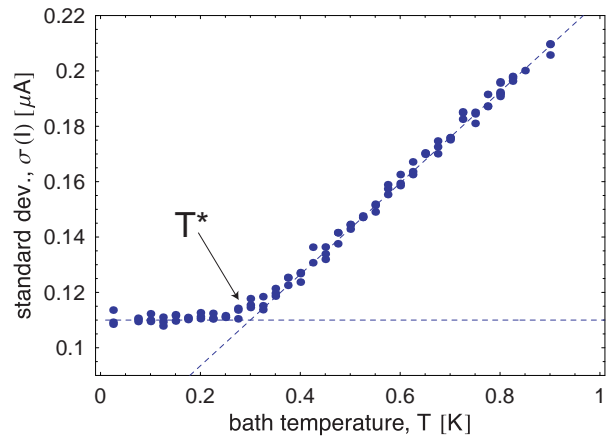}
	\caption{Standard deviation $\sigma_{I}$ of the $P(I)$ distributions 
	versus the bath temperature $T$. The cross-over temperature 
	$T^{\star}$ is indicated.}
	\label{fig:IStdevVsT}
\end{figure}

The standard deviation $\sigma_{I}$ of $P(I)$ is plotted versus $T$ in
Fig.~\ref{fig:IStdevVsT}.  In the temperature range between $1 \,
\rm{K}$ and approximately $300 \, \rm{mK}$, $\sigma_{I}$ decreases
with $T$, indicating the temperature dependent
thermal activation of the phase across the barrier.  At the
characteristic temperature $T^{\star} \approx 300 \, \rm{mK}$,
$\sigma_{I}$ saturates at approximately $115 \, \rm{nA}$ suggesting
that the escape of the phase is dominated by quantum tunneling
through the barrier.  In Sec.~\ref{sec:analysis}, the $P(I)$
distributions in both the thermal and the quantum regime are analyzed
quantitatively and compared with existing models.

\section{Experimental Resolution \label{sec:res}}

We have compared the experimental resolution of our switching current
measurements performed with the experimental setup described above to
the resolution of such measurements presented in literature.  Since
the limiting resolution of the measurement technique is not always
stated in existing publications, we have focused our comparison on the
minimum measured width of the switching current distribution mentioned
in the articles.  To allow for comparison among very widely varying
sample parameters, we express the width with respect to the critical
current of the measured junction.  As a measure of the width we have
considered the standard deviation $\sigma_{I}$ of the $P(I)$
distributions.  If the values of $\sigma_{I}$ were not available in
the reference, we evaluated the full width at half maximum of $P(I)$. 
In some of the evaluated references the effect of energy level
quantization is investigated.  This typically leads to features on the
distributions which are more narrow than $\sigma_{I}$.  In such cases
\cite{Silvestrini97,Martinis87} we have taken the width of
these smaller features.  If data of multiple samples were discussed in
a reference, the data of the sample with the smallest relative $P(I)$
width is considered.

In Tab.~\ref{tbl:resolution} the collected data are summarized.
\begin{table*}[tbp]
    \centering
    \caption{Experimental resolution in Josephson junction switching
    current experiments.  Listed are the reference from which data has
    been extracted (col 1), the focus subject of the respective reference
    (col 2), the critical current $I_{c}$ of the measured sample (col 3),
    the standard deviation $\sigma_{I}$ of the most narrow measured
    switching current distribution (col 3), $\sigma_{I}/I_{c}$ (col 4),
    the junction capacitance $C$ (col 5), the effective junction
    resistance $R$ (col 6), the cross-over temperature $T^{\star}$ (col
    7), the junction dimensions (col 8), and the materials used for
    junction fabrication (col 9). }
    \begin{tabular}{cccccccccc}
        \hline
	\hline
        Refs. &  
	focus &
	$I_{c}$ & 
	$\sigma_{I}$& 
	$\sigma_{I}/I_{c}$ & 
	$C$& 
	$R$ & 
	$T^{\star}$& 
	dimensions & 
	material \\ 
        &  
	&
	$\rm{[\mu A]}$ & 
	$\rm{[\mu A]}$ & 
	 & 
	[pF] & 
	$\rm{[\Omega]}$ & 
	[mK] & 
	[$\mu m \times \mu m $] & 
	\\ 
        \hline
        Washburn \textsl{et al.}\cite{Washburn84} & MQT & $57.4$ & -- & $0.9 \times 10^{-3}$ 
	& $0.67$  & -- & 500 & $1.2 \times 0.13$ & 
	Nb-Nb$_{2}$O$_{5}$-Nb \\
	 & MQT & $57.4$ & -- & $1.5 \times 10^{-3}$ 
	& $12.00$  & -- & 150 & $1.2 \times 0.13$ & 
	Nb-Nb$_{2}$O$_{5}$-Nb \\
 	Devoret, Martinis \textsl{et al.}\cite{Devoret85,Martinis87} & MQT, ELQ, TA & 9.489 & -- & $0.4 \times 10^{-3}$ &
 	6.35 & 190 & $37$ & $10 \times 10$ & Nb-NbO$_{x}$-PbIn \\
	Silvestrini \textsl{et al.}\cite{Silvestrini88b} & TA & $162$ & 0.245 & $1.5 \times 
 	10^{-3}$ & $500$ &  --  & -- & $100 \times 50$ & Nb-NbO$_{x}$-Pb  \\
 	Vion \textsl{et al.}\cite{Vion96} & TA & $0.040$ & -- & $0.4 \times 
 	10^{-3}$ & 0.00015 & -- & -- & -- & Al-AlO$_{x}$-Al \\
	Castellano \textsl{et al.}\cite{Castellano96b} & TA & $160$ & -- & $2 \times 
 	10^{-3}$ & 0.8 & -- & -- & $4 \times 4$ & Nb-Al-AlO$_{x}$-Nb \\
	Silvestrini \textsl{et al.}\cite{Silvestrini97} & TA & $47.0$ & -- & $7 \times 
 	10^{-3}$ & $5.5$ & $20 \times 10^{3}$ & -- & $10 \times 10$ & Nb-Al-AlO$_{x}$-Nb  \\
	                           & ELQ & $47.0$ & -- & $0.7 \times 
 	10^{-3}$ & $5.5$ & $20 \times 10^{3}$ & -- & $10 \times 10$ & Nb-Al-AlO$_{x}$-Nb  \\
	Ruggiero \textsl{et al.}\cite{Ruggiero98} & TA & $175$ & 0.300 & $2 \times 
 	10^{-3}$ & $2.5$ & -- & -- & $5 \times 5$ & Nb-Al-AlO$_{x}$-Nb  \\
        this work & MQT, TA & $315.0$ & $0.115$ & $0.4 \times 10^{-3}$ & 
        $1.00$  & $500$ & $300$ & $5 \times 5$ & Nb-Al-AlO$_{x}$-Nb\\
	\hline
	\hline
    \end{tabular}
    \label{tbl:resolution}
\end{table*}
First, we note that our sample has a high critical current and a
relatively small capacitance, leading to a comparatively high
cross-over temperature.  Only in
Refs.~\onlinecite{Washburn84,Jackel81} higher cross-over temperatures
have been reported.  Most of the experiments listed in
Tab.~\ref{tbl:resolution} reach measured relative widths
$\sigma_{I}/I_{c}$ down to a few $10^{-3}$.  Only in a few experiments
the smallest relative width is a few times $10^{-4}$.  The
measurements we present here show a resolution which is close to the
best measurements of this type performed so far, see last row of
Tab.~\ref{tbl:resolution}.  We stress that we have not performed any
digital filtering or data processing neither on the raw data nor on
the switching current distributions presented here.  In microwave
spectroscopy experiments performed on the same sample\cite{Wallraff02}
we have observed photon absorption peaks which showed even smaller
relative widths than the switching current distributions presented
here.  In test experiments we have measured current histograms of
relative width down to $1.5 \times 10^{-4}$, which is the limiting
experimental current resolution achievable in our setup at the present
time.  These tests have been performed with the junction biased in the
resistive state but otherwise with experimental conditions identical
to those used for performing switching current measurements.

We note that the \emph{instrumental resolution} of the switching
current measurements by a timing technique as implemented in this
setup is extremely high as in comparison to other approaches using
analog to digital converters \cite{Martinis87,Castellano96b, Ruggiero99a}
or time to amplitude converters \cite{Silvestrini88,Silvestrini88b}. 
However, we note that typically the \emph{experimental resolution} for
switching current measurements is not limited by the instrumental
resolution of the data acquisition hardware but rather by the residual
current and voltage noise in the junction bias electronics.  Thus the
most important aspect in a well designed measurement setup is the
reduction of the external interference by the use of appropriate
filtering techniques together with low noise analog electronics.

\section{Data Analysis
\label{sec:analysis}}

\subsection{Thermal Activation}
At temperatures $T > T^{\star}$ the escape of the phase is dominated
by thermal activation.  At a fixed bias current the activation rate is
given by Eq.~(\ref{eq:thermalrate}).  Using Eq.~(\ref{eq:RateCont}),
the experimental escape rate $\Gamma(I)$ is determined from the $P(I)$
data and compared with the predictions of Eq.~(\ref{eq:thermalrate}).

To perform this comparison we rewrite Eq.~(\ref{eq:thermalrate}) using
the known bias current dependences of the approximated barrier height
(\ref{eq:VoSJ}) and the small amplitude oscillation frequency 
(\ref{eq:smallFreqSJ}) in the following form
\begin{equation} 
    \left(\ln{\frac{2 \pi \Gamma(I)} {a_{t} \omega^{\phi}_{0}
    (I)}}\right)^{2/3} = \left(\frac{E_{J}}{k_{\rm{B}}
    T_{\rm{esc}}}\frac{4\sqrt{2}}{3}\right)^{2/3} \frac{1}{I_{c}}
    \left(I_{c} - I\right) \, ,
    \label{eq:sjjratenorm}
\end{equation}
where we have introduced the effective escape temperature
$T_{\rm{esc}}$ \cite{Devoret85}.  The left hand side of
Eq.~(\ref{eq:sjjratenorm}) is then calculated in a first iteration
using the experimental data $\Gamma(I)$ and $a_{t}
\omega^{\phi}_{0}(I)$ estimated approximating the initially unknown
value of $I_{c}$ by the maximum measured switching current
$I_{\rm{max}}$ and using $C = 1.61 \, \rm{pF}$ determined from
independent microwave resonance activation measurements on the same
sample\cite{Wallraff02}.  Because the right hand side of
Eq.~(\ref{eq:sjjratenorm}) is a linear function of the bias current
$I$, both the critical current $I_{c}$ in absence of fluctuations and
the effective temperature $T_{\rm{esc}}$ of the thermal activation
process can be determined from a linear fit of the experimental data
to the right hand side of Eq.~(\ref{eq:sjjratenorm}).  Accordingly, we
find
\begin{eqnarray}
    I_{c} & = & \frac{c_{\rm{const}}}{c_{\rm{lin}} } \, ,
    \label{eq:IcSJ} \\
    T_{\rm{esc}} & = & \frac{1}{k_{\rm{B}}}\frac{\Phi_{0}}{2 \pi} 
    \frac{4 \sqrt{2}}{3}
     \frac{1}{c_{\rm{const}}^{1/2} c_{\rm{lin}}}  \, ,
    \label{eq:TescSJ} 
\end{eqnarray}
where $c_{\rm{const}}$ and $c_{\rm{lin}}$ are the two fitting
parameters.  The values of $I_{c}$ and $T_{\rm{esc}}$ are then found
with high accuracy by iteratively repeating the fitting procedure with
the value of $I_{c}$ found in the previous iteration
\cite{Castellano96b}.  This procedure is converging quickly due to the
logarithmic dependence of the left hand side of
Eq.~(\ref{eq:sjjratenorm}) on $I_{c}$.

\subsubsection{The thermal prefactor in the extremely low damping limit
\label{sec:thermalprefactor}}
We have extracted the effective escape temperature $T_{\rm{esc}}$ from
the experimental data according to the procedure described above
taking into account the thermal prefactor $a_{t}$.  In the simple
transition state theory, the thermal prefactor is unity and the escape
rate is only determined by the fraction of particles with energy
larger than the barrier height.  In the high as well as in the low
damping regimes the prefactor deviates substantially from unity
\cite{Weiss99}.

The exact form of the prefactor is found from a analysis of a
classical Langevin equation for the motion of the particle according
to Eq.~(\ref{eq:smalljunctiondynamicsparticle}) in the presence of
fluctuations (see e.g. Refs.~\onlinecite{Weiss99,Haenggi90b}).  In the
moderate to strong damping limit the value of $a_{t}$
depends predominantly on the damping and the small oscillation
frequency in the junction.  In the very strong damping limit, the
inertia of the particle can be neglected and its motion becomes
diffusive \cite{Vion96,Martinis89}.  In the low to extremely low
damping regime, however, the distribution of the particles in the well
deviates from thermal equilibrium.  The thermal escape across the
barrier depletes the population within a range of $k_{\rm{B}} T$ to
the top of the barrier.  Therefore, $a_{t}$ becomes dependent on the
barrier height and on temperature.  For simplicity, we use an
approximate analytical expression for the thermal prefactor in the
extremely low to low damping regime, relevant for the experimental
data presented here, as calculated in Ref.~\onlinecite{Buttiker83}
\begin{equation}
    a_{t} = \frac{4 a_{0}}{\left(\sqrt{1+\frac{a_{0} Q k_{\rm{B}}
    T}{1.8 \, U^{\phi}_{0}}}+1\right)^{2}} \, ,
    \label{eq:Buttikerprefactor}
\end{equation}
where $a_{0}$ is a numerical constant close to unity.  An exact
calculation of $a_{t}$ valid in all damping regimes can be found in
Ref.~\onlinecite{Haenggi90b}.

\begin{figure}[tb]
 	\centering
	\includegraphics[width = 1.0 \columnwidth]{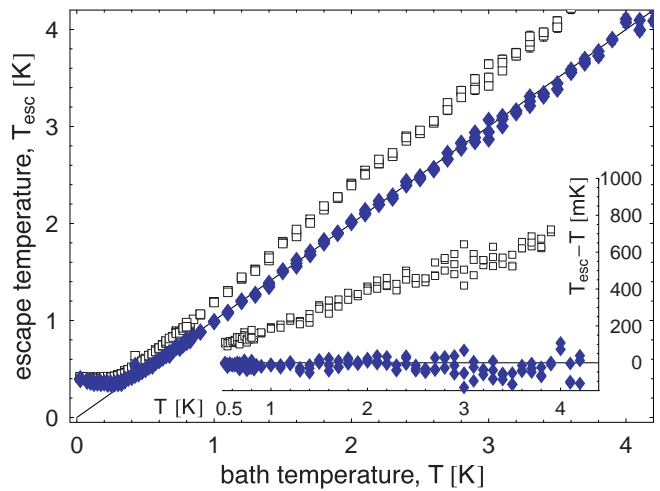}
	\caption{Escape temperature $T_{\rm{esc}}$ extracted from experimental
	data versus bath temperature $T$ for $a_{t} = 1$ (open symbols) and
	$a_{t}$ according to Eq.~(\ref{eq:Buttikerprefactor}) (solid symbols). 
	At high $T$ the statistical error in $T_{\rm{esc}}$ extracted from the
	fitting procedure is approximately given by the symbol size, at low $T$
	it is substantially less than the symbol size.  The solid line
	corresponds to $T = T_{\rm{esc}}$. The inset shows $T_{\rm{esc}} - T$.}
	\label{fig:Tesc}
\end{figure}

In a first step, $T_{\rm{esc}}$ has been evaluated for $a_{t} = 1$. 
Under this assumption the escape temperatures extracted from the data
deviates more than $100 \, \rm{mK}$ from the bath temperature at $T >
T^{\star}$, see open symbols in Fig.~\ref{fig:Tesc}.
The difference $T_{\rm{esc}} - T$ even increases with increasing
temperature, as shown in the inset of Fig.~\ref{fig:Tesc}.  This
deviation is too large to be due to an inaccurate estimate of the
junction capacitance or errors in the calibration of the current
measurement, thus, indicating the importance of the thermal prefactor.

In a second step, we used the thermal prefactor given by the
expression (\ref{eq:Buttikerprefactor}) in the calculation of the left
hand side of Eq.~(\ref{eq:sjjratenorm}) in our data analysis scheme. 
We used the junction resistance $R$ as the adjustable parameter
determining the junction quality factor $Q$.  We found good agreement
between the extracted $T_{\rm{esc}}$ and the bath temperature $T$ for
$R = 500 \, \Omega$ neglecting its temperature dependence, see solid
symbols in Fig.~\ref{fig:Tesc}.  With the low damping prefactor
(\ref{eq:Buttikerprefactor}), the difference $\left| T_{\rm{esc}}-T
\right|$ is less than $75 \, \rm{mK}$ in the temperature range above
$T^{\star}$, as shown by the solid symbols in the inset of
Fig.~\ref{fig:Tesc}.  The effective resistance used for the fit is in
good agreement with the sub-gap resistance determined from the dc
current voltage characteristic of the junction.

\begin{figure}[!t]
 	\centering
	\includegraphics[width = 1.0 \columnwidth]{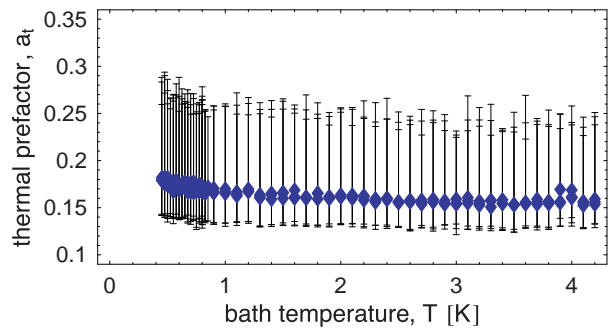}
	\caption{Thermal prefactor $a_{t}$
	[Eq.~(\ref{eq:Buttikerprefactor})] calculated at the most probable
	switching current $I_{p}$ versus bath temperature $T$.  The error bars
	indicate the value of $a_{t}$ at the minimum $I_{\rm{min}}$ and 
	maximum switching current $I_{\rm{max}}$ within each $P(I)$ 
	distribution.
	}
	\label{fig:ThermalPrefactor}
\end{figure}

In Fig.~\ref{fig:ThermalPrefactor}, the thermal prefactor $a_{t}$
evaluated at the most probable switching current $I_{p}$ is plotted
versus temperature for $T>T^{\star}$.  Independently of the bath
temperature, the mean value of $a_{t}$ is approximately $0.17$ .  This
weak temperature dependence can be attributed to the almost constant
ratio of $k_{\rm{B}} T / U_{0}^{\phi}(I_{p})$ for the switching data
measured at different temperatures, see
Eq.~(\ref{eq:Buttikerprefactor}).  At a fixed bath temperature, the
bias current variation within a single $P(I)$ distribution results in
a noticeable variation of $a_{t}$.  The value of $a_{t}$ at the
minimum and maximum switching currents $I_{\rm{min}}$ and
$I_{\rm{max}}$ is indicated in Fig.~\ref{fig:ThermalPrefactor} by
error bars.

Thus, the experimental data is explained with good accuracy by the classical
model of the thermal activation of the phase over a potential barrier
in the presence of extremely small damping.  The quality factor used
for the extraction of the effective escape temperature is in
agreement with the value determined from the sub gap resistance of
the junction.

\subsubsection{Thermal activation in the presence of energy levels}

In thermal activation theory, it is assumed that the particle energies
in the well are continuously distributed.  For the sample considered
here, however, the quantization of the energy of the small amplitude
oscillations of the phase may be relevant.  In fact, we found direct
evidence for the energy level quantization of the phase in this sample
by using microwave spectroscopy \cite{Wallraff02}.

\begin{figure}[tb]
 	\centering
	\includegraphics[width = 0.45 \columnwidth]{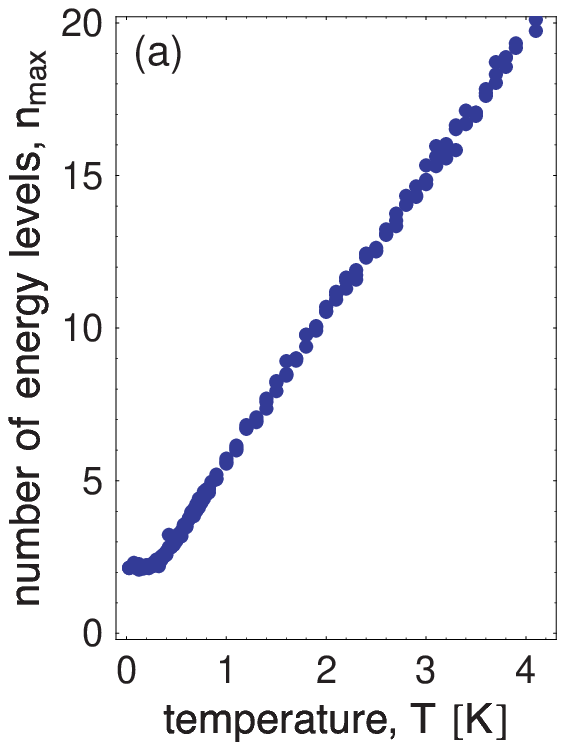} 
	\includegraphics[width = 0.45 \columnwidth]{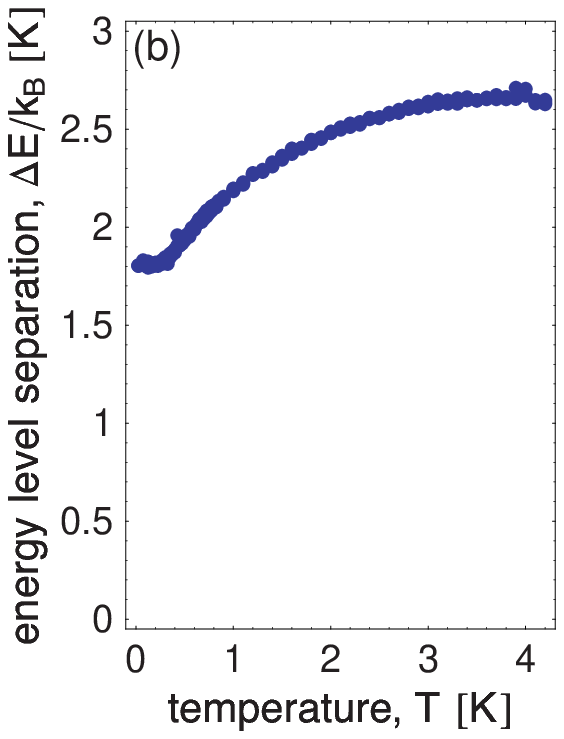}
	\caption{a) Approximate number of quantized energy levels
	$n_{\rm{max}}$ in the well at the most probable switching current
	$I_{p}$ versus the bath temperature $T$.  b) Harmonic approximation of
	energy level separation $\Delta E / k_{\rm{B}}$ at $I_{p}$ in Kelvin
	versus $T$.}
	\label{fig:levels}
\end{figure}

To address this issue we calculate the approximate number of
levels in the well at a given bias current as
\begin{equation}
    n_{\rm{max}} = \frac{U_{0}^{\phi}}{\hbar 
    \omega_{0}^{\phi}} \, ,
    \label{eq:numlevels}
\end{equation}
neglecting the anharmonicity of the well.  Eq.~(\ref{eq:numlevels})
has been evaluated at the most probable switching current $I_{p}$ at
each temperature, see Fig.~\ref{fig:levels}a.  In the quantum
tunneling regime at $T < T^{\star}$, we find that there are only a few
levels in the well ($n_{\rm{max}} \leq 3$).  The number of levels in
the well at the relevant switching currents increases approximately
linearly with temperature from about 5 levels at $1 \, \rm{K}$ to
about $20$ levels at $4 \, \rm{K}$.  Additionally we have evaluated
the level separation $\Delta E = \hbar \omega_{0}^{\phi}(I)$ in
harmonic approximation at the most probable switching current $I_{p}$,
see Fig.~\ref{fig:levels}b.  At temperatures below $1 \, \rm{K}$ the
level separation is much larger than $k_{\rm{B}}T$ and at temperatures
above $1 \, \rm{K}$ it is comparable with $k_{\rm{B}}T$. 
These facts indicate that the energy level structure may be relevant
for the process of the escape of the phase even at temperatures $T >
T^{\star}$.

To analyze the effect of the level quantization on the escape of
the phase at high temperatures we have made use of the Larkin
Ovchinnikov theory \cite{Larkin86}.  In the framework of their model
the escape of the phase is considered in terms of tunneling out of the
well from individual energy levels.  The overall escape rate
$\Gamma_{LO}(I)$ depends on the population of different levels, which 
is determined by the coupling to the thermal bath.  To apply the
Larkin Ovchinnikov theory, we have calculated the energy level
structure of the phase in the potential in dependence on the bias
current $I$.  The matrix elements for transitions between individual
levels have been evaluated and the resulting escape rate has been
calculated by solving a master equation for the dynamics of the
system.  From the escape rate $\Gamma_{LO}(I)$ we have calculated
$P(I)$ distributions in a wide temperature range.

The calculated $P(I)$ distributions at temperatures well below
$T^{\star}$, when essentially only the ground state is populated, were
in good agreement with the distributions calculated using the WKB
approximation for the tunneling from the ground state.  At
temperatures above $T^{\star}$, the width of the distribution
calculated using Larkin Ovchinnikov theory increased with temperature
as expected.  We have evaluated these calculated data in the same way
as the experimental data, see Sec~\ref{sec:thermalprefactor}.  We
found that the calculated histograms are consistent with the classical
prediction for a low damping thermal prefactor.  Thus in the high
temperature limit the results of the Larkin Ovchinnikov theory are in
agreement with the predictions of purely classical thermal activation
theory.
More details on this analysis will be presented elsewhere
\cite{Duty02}.

\subsection{Quantum Tunneling}

At temperatures below the crossover temperature\cite{Grabert84}
\begin{equation}
    T^{\star} = \frac{\hbar \omega^{\phi}_{0}}{2 \pi k_{\rm{B}}}
    \left[
    \left(1+\left(\frac{1}{2Q}\right)^{2}\right)^{1/2}-\frac{1}{2Q}
    \right]
    \label{eq:corssoverT}
\end{equation}
the escape of the phase from the potential well is dominated by
quantum tunneling through the barrier.  According to
Eq.~(\ref{eq:corssoverT}), the crossover temperature $T^{\star}$ for
this sample at the most probable switching current $I_{p}$ is
predicted to be approximately $282 \, \rm{mK}$, which is in good
agreement with our experimental findings, see
Fig.~\ref{fig:IStdevVsT}.  Since the quality factor of this sample is
bounded by $Q > 100$, the reduction of $T^{\star}$ in comparison to
the ideal case of $Q = \infty$ is less than one percent.  Accordingly,
the reduction of the tunneling rate due to dissipation is small.

\begin{figure}[tb]
 	\centering
	\includegraphics[width = 0.95 \columnwidth]{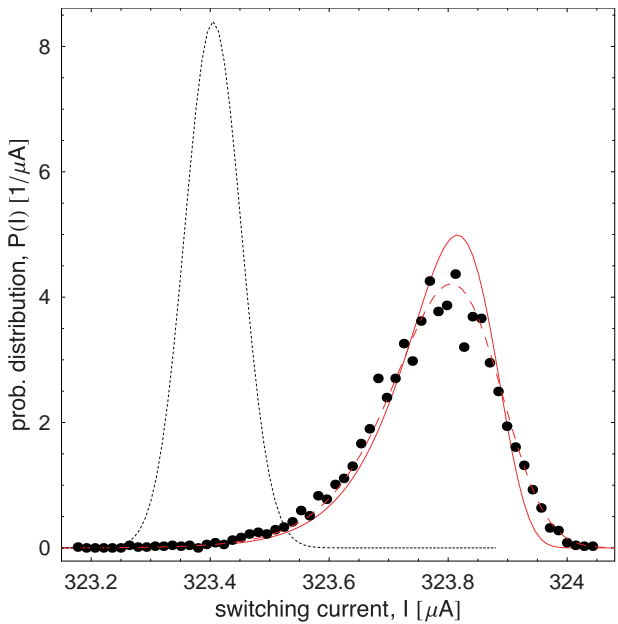}
	\caption{Measured switching current distribution $P(I)$ at $T = 25
	\rm{mK}$ (solid symbols).  The solid line is the predicted switching
	current distribution $P_{q}(I)$ in the quantum tunneling regime for
	$I_{c} = 325.05 \, \rm{\mu A}$ and $C = 1.61 \, \rm{pF}$.  The
	$P_{q}(I)$ distribution broadened by residual experimental noise
	modeled by gaussian distribution with $\sigma_{I} = 47.5 \, 
	\rm{nA}$ (dotted line) is shown by the dashed line.}
	\label{fig:QuantumPfit}
\end{figure}

Furthermore, we have compared the measured $P(I)$ distributions for $T
\ll T^{\star}$ to the predicted distributions calculated using the
quantum tunneling rate~(\ref{eq:WKBrate}).  As an example, the data
for $T = 25 \, \rm{mK}$ are shown in Fig.~\ref{fig:QuantumPfit} by
solid symbols.
The quantum tunneling distribution $P_{q}(I)$ calculated for $I_{c} =
325.05 \, \rm{\mu A}$ and $C = 1.61 \, \rm{pF}$ is shown by the solid
line.  The critical current $I_{c}$ was extracted from fits to the
data in the thermal regime.  As it has been already said above, the
capacitance $C$ was determined with high accuracy in independent
spectroscopic measurements of the energy level structure of the same
sample \cite{Wallraff02}.  The most probable switching current of the
calculated distribution is in good agreement with experimental data. 
However, the width of the measured distribution is slightly bigger
than the predicted one, which is most likely due to the finite
residual electromagnetic interference in the measurement setup.  To
quantify this effect, we have calculated the convolution of the
predicted quantum tunneling distribution $P_{q}(I)$ with a gaussian
current distribution with a standard deviation of $\sigma_{I} = 47.5
\, \rm{nA}$, shown by the dotted line centered about $I = 323.4 \,
\rm{\mu A}$.  The resulting distribution, shown by the dashed line in
Fig.~\ref{fig:QuantumPfit}, accurately fits to the experimental data.

\section{Conclusions \label{sec:conclusions}}

We have developed an experimental setup for the investigation of the
escape of the phase in a current-biased Josephson tunnel junction. 
Using a timing technique we are able to achieve a high instrumental
resolution in the switching current measurement of a Josephson
junction.  The relative resolution of the switching current
measurements demonstrated in this paper is better than $3.7 \times
10^{-4}$.  This distribution width $\sigma_{I}/I_{c}$ is comparable to
the highest resolution measurements of this type performed so far. 
Using the described technique we have investigated the escape of phase
in a high-quality $5 \times 5 \, \rm{\mu m^{2}}$ Nb-Al/AlO$_{x}$-Nb
tunnel junction with a relatively large critical current of $I_{c}
\approx 325 \, \rm{\mu A}$.  At temperatures above the crossover
temperature of $T^{\star} \approx 300 \, \rm{mK}$ the escape is
dominated by thermal activation.  Due to the high quality factor of
the junction ($Q > 100$) the thermal activation is observed in the
limit of extremely low damping.  The thermal prefactor $a_{t}$
extracted from the experimental data is varying between $0.1$ to $0.3$
depending on bias current and temperature.  At temperatures below
$T^{\star}$, the value of which agrees well with the theoretical
predictions, the escape of the phase by quantum mechanical tunneling
is observed.  Due to the high $Q$-factor of the junction, the
tunneling rate is only weakly suppressed by dissipation.

Recent developments in the field of quantum information processing
have strongly increased the interest in research on the quantum
properties of Josephson junction systems
\cite{Nakamura99,Friedman00,vanderWal00}.  Using the measurement
technique and setup described in this paper the quantum mechanical
properties of current biased Josephson junctions can be investigated
with high accuracy.  In particular, we are currently applying this
technique for the experimental investigation of the theoretically
predicted\cite{Kato96,Shnirman97b} but not yet observed quantum
properties of a single vortex in a long Josephson junction
\cite{Wallraff00phdbook}.

\begin{acknowledgments}
    We would like to thank A.~Kemp, S.~Lebeda, G.~Logvenov, K.~Urlichs and
    M.~Schuster for their technical help.  We are indebted to M.~H.
    Devoret, M.~Fistul and V.~Kurin for useful discussions and to C. van der
    Wal, R. Schouten for for sharing their experience in low temperature
    transport measurements with us.  We acknowledge the partial financial
    support of this project by the Deutsche Forschungsgemeinschaft (DFG)
    and by D-Wave Systems Inc..
\end{acknowledgments}

\end{document}